\documentclass[longauth]{aa} 
\usepackage{graphicx}
\usepackage{txfonts}

\newcommand{\meth}{CH$_{\mathrm 4}$}
\newcommand{\Ms}{CH$_{\mathrm 4}$s}
\newcommand{\Ml}{CH$_{\mathrm 4}$l}
\newcommand{\Msl}{\Ms$-$\Ml}
\begin{document}
   \title{Two T dwarfs from the UKIDSS Early Data Release}

   \author{T.R Kendall
          \inst{1}
          \and
          M. Tamura
          \inst{2}
          \and
          C.G. Tinney
          \inst{3} 
          \and
          E.L. Mart\'{i}n
          \inst{4,5}          
          \and
          M. Ishii
          \inst{6}
          \and
          D.J. Pinfield
          \inst{1}
          \and 
          P.W. Lucas
          \inst{1}
          \and
          H.R.A. Jones
          \inst{1}
          \and
          S.K. Leggett
          \inst{7}
          \and
          S. Dye
          \inst{8}
          \and
          P.C. Hewett
          \inst{9} 
          \and
          F. Allard
          \inst{10}
          \and
          I. Baraffe
          \inst{10}
          \and          
          D. Barrado y Navascu\'{e}s
          \inst{11}
          \and
          G. Carraro
          \inst{12}
          \and
          S.L. Casewell
          \inst{13}
          \and
          G. Chabrier
          \inst{10}
          \and
          R.J. Chappelle
          \inst{14}
          \and
          F. Clarke
          \inst{15}
          \and
          A. Day-Jones
          \inst{1}
          \and
          N. Deacon
          \inst{16}
          \and
          P.D. Dobbie
          \inst{13}
          \and
          S. Folkes
          \inst{1}
          \and
          N.C. Hambly
          \inst{16}
          \and
          S.T. Hodgkin
          \inst{9}
          \and
          T. Nakajima
          \inst{2}
          \and
          R.F. Jameson
          \inst{13}
          \and
          N. Lodieu
          \inst{13}
          \and
          A. Magazz\`{u}
          \inst{17}
          \and
          M.J. McCaughrean
          \inst{18}
          \and
          Y.V. Pavlenko
          \inst{19}
          \and
          N. Tadashi
          \inst{2}
          \and
          M.R. Zapatero Osorio
          \inst{4}
}

   \offprints{T.R. Kendall}

   \institute{Centre for Astrophysics Research, Science \& Technology Research Institute,
              University of Hertfordshire, College Lane, Hatfield AL10 9AB, U.K.
              \email{trkendall@googlemail.com, d.j.pinfield@herts.ac.uk}
         \and
             National Astronomical Observatory of Japan, 2-21-1 Osawa, Mitaka, Tokyo 181-8588 Japan
         \and
             Anglo-Australian Observatory, PO Box 296, Epping 1710, Australia  
         \and
             Instituto de Astrof\'{i}sica de Canarias, 38200 La Laguna, Spain
         \and
             University of Central Florida, Physics Dept., PO Box 162385, Orlando, FL 32816-2385, USA
         \and
             Subaru Telescope, 650 North A'ohoku Place, Hilo, Hi 96720, USA
         \and
             Gemini Observatory, 670 North A'oholu Place, Hilo, Hawai'i 96720 
         \and
             School of Physics \& Astronomy, Cardiff University, 5, The Parade, Cardiff, CF24 3AA, Wales, UK
         \and
             Inst. of Astronomy, Madingley Rd., Cambridge CB3 0HA, U.K.
         \and
             C.R.A.L. (UMR 5574 CNRS) Ecole Normale Sup\'{e}riure, 69364 Lyon Cedex 07, France          
         \and
             Laboratorio de Astrof\'{i}sica Espacial y F\'{i}sica Fundamental, INTA, PO Box 50727, 28080 Madrid, Spain 
         \and
             Dipartmento di Astronomia, Universit\'{a} di Padova, Vicolo Osservatorio 5, I-35122 Padova, Italy
         \and
             Dept. of Physics \& Astronomy, University of Leicester, LE1 7RH, U.K.
         \and
             Astronomical Institute, Academy of Sciences of the Czech Republic, Bocni II/1401a, 141 31 Prague, Czech Republic
         \and
             European Southern Observatory, Alonso de Cordova 3107, Casilla 19001 Santiago 19, Chile
         \and
             Scottish Universities Physics Alliance, Institute for Astronomy, School of Physics, University of Edinburgh,
             Royal Observatory, Blackford Hill, Edinburgh, EH9 3HJ, U.K.
         \and
             Fundaci\'{o}n Galileo Galilei-INAF, Apartado 565, 38700 Santa Cruz de La Palma, Spain
         \and
             University of Exeter, School of Physics, Stocker Road, Exeter, EX4 4QL, U.K.
         \and
             Main Astronomical Observatory, National Academy of Sciences, Zabolotnoho, Kyiv-127, 03680 Ukraine
}

   \date{Received; accepted}

  \abstract
{We report on the first ultracool dwarf discoveries from the UKIRT Infrared 
Deep Sky Survey (UKIDSS) Large Area Survey Early Data Release (LAS EDR), in 
particular the discovery of T dwarfs which are fainter and more distant than 
those found using the 2MASS and SDSS surveys.}
{We aim to show that our methodologies for searching the $\sim$\,27\,deg$^2$ 
of the LAS EDR are successful for finding both L and T dwarfs $via$ cross-correlation 
with the Sloan Digital Sky Survey (SDSS) DR4 release. While the area searched 
so far is small, the numbers of objects found shows great promise for near-future 
releases of the LAS and great potential for finding large numbers of such dwarfs.}
{Ultracool dwarfs are selected by combinations of their $YJH(K)$ UKIDSS colours 
and SDSS DR4 $z-J$ and $i-z$ colours, or, lower limits on these red optical/infrared 
colours in the case of DR4 dropouts. After passing visual inspection tests, 
candidates have been followed up by methane imaging and spectroscopy at 4m and 
8m-class facilities.}
{Our main result is the discovery following CH$_4$ imaging and spectroscopy of a 
T4.5~dwarf, ULAS\,J\,1452+0655, lying $\sim$\,80\,pc distant. A further T 
dwarf candidate, ULAS\,J\,1301+0023, has very similar CH$_4$ colours but has not 
yet been confirmed spectroscopically. We also report on the identification of a 
brighter L0 dwarf, and on the selection of a list of LAS objects designed to probe 
for T-like dwarfs to the survey $J$-band limit.}
{Our findings indicate that the combination of the UKIDSS LAS and SDSS surveys 
provide an excellent tool for identifying L and T dwarfs down to much fainter 
limits than previously possible. Our discovery of one confirmed and one probable 
T dwarf in the EDR is consistent with expectations from the previously measured 
T dwarf density on the sky.}


   \maketitle
%

\section{Introduction}

The goals of the UKIRT Infrared Deep Sky Survey (UKIDSS) have been described by \cite{law06} and a 
full technical description of the Early Data Release (EDR) is given by \cite{dye06}. One of the prime 
science drivers for the UKIDSS Large Area Survey (LAS) is to search for large numbers of brown dwarfs, 
including those cooler than any known hitherto, i.e. with effective temperatures less than the latest 
T dwarfs ($\sim$ 700--800K). This is made possible by the depth and coverage of the survey as well as 
the use of the Y filter, covering 0.97 to 1.07\,$\mu$m. This filter is designed to allow such objects 
to be distinguished from main-sequence stars and high-$z$ quasars (\cite{leg05}). The UKIDSS $ZYJHK$ 
system is described by \cite{hew06} and simulations of the projected results of the LAS concerning 
ultracool dwarfs described by \cite{dea06}.

\begin{table*}
\caption{Five SDSS dropout T dwarf candidates and a confirmed L dwarf followed up by methane imaging 
and/or spectroscopy. The first five objects listed are from the seven T dwarf candidates obtained 
in the $YJH$ search, where two have detected methane one of which is confirmed spectroscopically, 
and three are methane non detections. The Last object listed is the spectroscopically confirmed L 
dwarf from the first $YJHK$ search. Uncertainties on $YJH$ magnitudes are taken from the LAS EDR 
merged catalogue, as are $Y-J$ and $J-H$ colours. The penultimate column gives the CH$_4$s\,$-$\,CH$_4$l 
index from AAT/IRIS2 imaging, where applicable.}
\label{t1}
\centering
\begin{tabular}{c c c c c c c c}     
\hline\hline       
Name (RA  dec) & $Y$ & $J$ & $H$ & $Y-J$ & $J-H$ & CH$_4$ & comment \\  
\hline                    
ULAS\,J145243.59+065542.9 & 19.77\,$\pm$\,0.14 & 18.66\,$\pm$\,0.08 & 18.45\,$\pm$\,0.15 & 1.11 & 
0.21 & $-$0.17\,$\pm$\,0.13 & T4.5 (Subaru spectroscopy)\\   
ULAS\,J130150.35+002314.8 & 20.31\,$\pm$\,0.18 & 19.24\,$\pm$\,0.13 & 18.92\,$\pm$\,0.20 & 1.07 & 
0.32 & $-$0.23\,$\pm$\,0.15 & CH$_4$ detection \\
\hline  
ULAS\,J130519.86+001402.9 & 20.61\,$\pm$\,0.23 & 19.39\,$\pm$\,0.15 & 19.28\,$\pm$\,0.27 & 1.23 & 
0.11 & +0.19\,$\pm$\,0.14 & CH$_4$ non-detection \\
ULAS\,J130153.49+002044.9 & 20.74\,$\pm$\,0.26 & 19.38\,$\pm$\,0.15 & 18.91\,$\pm$\,0.20 & 1.36 & 
0.47 & +0.17\,$\pm$\,0.13 & CH$_4$ non-detection \\
ULAS\,J152347.60+052849.4 & 19.61\,$\pm$\,0.11 & 18.68\,$\pm$\,0.08 & 18.38\,$\pm$\,0.10 & 0.93 & 
0.30 & - & field (Subaru spectroscopy)\\
\hline
ULAS\,J153108.89+060111.1 & 16.83\,$\pm$\,0.01 & 15.81\,$\pm$\,0.01 & 15.07\,$\pm$\,0.01 & 1.02 & 
0.74 & - & L0 (TNG spectroscopy) \\
\hline                  
\end{tabular}
\end{table*}

In this paper, we describe early systematic efforts to probe the ultracool dwarf population observed by 
the LAS EDR, covering $\sim$\,27\,deg$^2$ on the sky in all four filters $YJHK$ (c.f. 4000\,deg$^2$ for 
the complete LAS, to be observed over a seven year duration). We present our successful search methodologies 
in Sect. 2 and detail results from imaging and spectroscopic follow-up at the AAT, TNG and Subaru in Sect. 3, 
before summarising in Sect. 4.

\section{Catalogue search methodology}

\begin{figure}
   \centering
   \includegraphics[width=9cm]{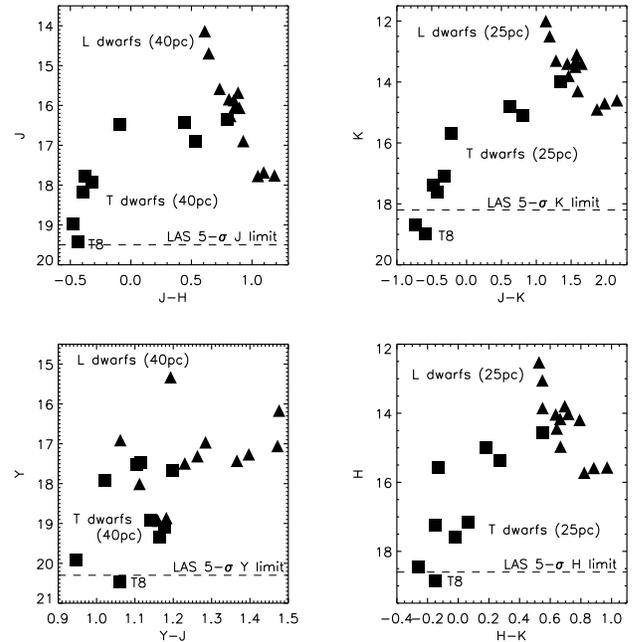}
      \caption{Example illustrations of the UKIDSS LAS sensitivity to ultracool dwarfs, compared to empirical 
colour data derived from \cite{hew06} for L1 -- T8 dwarfs (triangles and squares for L and T types respectively) 
with known parallaxes and absolute magnitude, and shifted to the distances indicated. The horizontal lines near 
the bottom of the plots indicate the LAS sensitivity limits. At $K$, late T dwarfs at 25\,pc would be missed. 
In order to be sensitive to these, more numerous, distant T dwarfs, we thus chose to require $K$-band non-detections 
in our T dwarf search. While this approach could potentially miss some nearer ($<$17pc), brighter objects, we 
do not expect a significant number in the 27~deg$^2$ of the EDR. We thus chose to aim for the more numerous 
distant population. It can be seen that our search will be $H$-band limited. Searches for $YJ$-only detections 
can thus potentially probe similar objects at much greater distances (m$-$M\,=\,3 is shown).}
   \label{f1}
\end{figure}

The limiting magnitudes of the LAS are $Y$\,=\,20.3, $J$\,=\,19.5, $H$\,=\,18.6 and $K$\,=\,18.2, 
defined on the Vega system for a S/N\,=\,5 point source detection in a 2\arcsec~ diameter aperture; 
\cite{dye06}. The precise $\alpha$, $\delta$ coverage of the 27$^2$degs of the EDR is also described 
by \cite{dye06} and is contained within the Sloan Digital Sky Survey (SDSS) DR4 release. We have 
performed three different source searches involving cross-correlation of the LAS EDR with SDSS DR4. 
The first search is designed to seek L dwarfs, with LAS $K$-band detections, and the second to detect 
T dwarfs without $K$-band detections, in an $H$-band limited search (see Fig.\,\ref{f1}). The third 
search is designed to find blue late T dwarf-like objects with $YJ$-only detections.

The first search was designed to yield L and early T candidates, and used the following criteria: $Y-J$ 
in the range 0.9 to 1.7, $J-H$ in the range 0.5 to 1.5 and $H-K$ in the range 0.0 to 1.3. All objects 
were required to be $YJHK$ detections with pStar\,$>$\,0.9 and mergedClass\,=\,-1
\footnote{see \tt http://surveys.roe.ac.uk/wsa/ {\rm pStar is the probability of the detection 
being a point source (a value between 0 and 1). mergedClass is a Class flag from all available 
measurements: 1 galaxy, 0 noise, $-$1 stellar, $-$2 probable stellar, $-$3 probable galaxy, 9 saturated.}}. 
These colour constraints produced 114 candidates which were inspected on UKIDSS images. None were found to 
be cross-talk artifacts and all were within the SDSS DR4 footprint. SDSS DR4 cross-correlation was performed 
with a 1.2\arcsec~ radius search for nearby matches, followed by a 0.3\arcmin~ radius search for possible 
high proper motion counterparts, at some separation from the UKIDSS coordinates.

This process produced 88 nearby counterparts of which 30 conformed to the additional constraints of 
$z-J$\,$>$\,2 and $i-z$\,$>$\, 1.5, corresponding to an M8 dwarf or later. All SDSS (AB) magnitudes are 
point-spread-function fitted magnitudes, and for this sample of 30, one-sigma errors on $z$ are $<$\,0.3 
and on $i$, $<$\,0.35. The $z-J$ colour for these candidates ranges up to ~3, corresponding approximately 
to the L/T boundary (\cite{kna04}). Six additional candidates were also revealed by the 0.3\arcmin~ search 
for high proper motion counterparts, with DR4 catalogue $iz$ magnitudes conforming to the same ultracool 
criteria as before. Also, 3 LAS sources contained in the DR4 footprint had no catalogue counterpart. Two 
of the SDSS images of these candidates may show a very faint detection at $z$, although one might be a 
pair of hot pixels. An examination of SDSS $z$-band uncertainties for neighboring objects shows that a 
typical one-sigma error on $z$\,$\sim$\,0.4 is reached at $z$\,$\sim$\,21.5. If this is taken as an 
approximate limit for a real $z$-band detection, then the three dropouts here have $z-J$\,$\ga$\,3, 2.8 
and 2.3 respectively, corresponding to expected spectral types of T0, L6 and L0. In summary, there are 
39 ultracool dwarf candidates identified by this search.

The second search was for later-type T dwarfs. The latest T dwarfs have $J-K<$0, and if one requires a 
K-band detection for such objects, the LAS K-band sensitivity will unnecessarily limit the distance and 
volume probed (see Fig 1). The T dwarf search was thus made requiring $K$-band non-detections. While 
this approach could potentially miss some nearer ($<$17pc), brighter objects, we do not expect a 
significant number in the 27~deg$^2$ of the EDR, and we thus chose to aim for the more numerous distant 
population. Colour criteria used were $Y-J$\,$>$\,0.9 and $J-H$\,$<$\,0.5 together with a likelihood of 
stellarity requirement that mergedClass\,=\,$-$1 or $-$2. Note that the first two searches cover adjacent 
but not overlapping regions in the $Y-J,J-H$ colour-colour diagram with the boundary at $J-H$\,=\,0.5. 
SDSS DR4 cross correlation was performed as for the first search, requiring either a red $z-J$\,$>$\,2 
colour or an optical non detection. Finally, image inspection was carried out to identify cross-talk and 
other spurious sources. The search resulted in seven T dwarf candidates.

All seven candidates are SDSS dropouts. LAS YJH zero-points were checked in the fits image headers, 
to ensure that they were not significantly different to other survey zero points derived from observations 
made on the same night (if they were, this could indicate a problem with the pipeline calibration). 
None of the relevant zero points were found to be significantly different from their associated nightly 
averages, with values typically $Y$\,=\,22.7, $H$\,=\,24.5 and $J$\,=\,24.7. Ellipticities measured for 
the detected sources, and given in the LAS merged catalogue, are maximum $\sim$\,0.3 at $Y$ and typically 
$\sim$\,0.15--0.2 at $J$ and $H$, quantifying further the maximum allowed deviation from a point source. 
One-sigma errors on $YJH$ are typically $\sim$\,0.2, 0.15 and 0.2 respectively. All the candidates except 
one are faint in the $H$-band ($\sim$\,19), and thus have $J-H$ colours indicative of mid-T dwarfs. The 
one exception is the reddest in both $Y-J$ and $J-H$ and might be more likely a late L or early T dwarf.

Thirdly, a search for $YJ$-only detections has been carried out, motivated by the wish to 
identify late T dwarfs with blue $J-H$ and $J-K$ colours, and possibly even cooler bluer objects. 
The benefits of this approach are made clear when one considers the $J$-band depth that can be 
probed. For an object with $J-H$\,=\,$-$1, one would only reach $J$\,=\,17.6 in the EDR if an 
$H$-band detection were required. Even for $J-H$\,=\,0, one would still only reach $J$\,=\,18.6. 
T dwarfs have $Y-J$\,$\sim$\,1 (Fig.\,\ref{f1}) so a $YJ$-only search should be capable of probing 
for such objects to the $J$-band limit, resulting in a several-fold increase in search volume.

Candidates were selected to have $Y-J$\,$\geq$\,0.8 and SNR\,$>$\,5 detections (errors on $Y$ and 
$J$\,$\leq$\,0.2). A constraint $J$\,$\leq$\,19.5 was employed, together with mergedClass\,=\,$-$1 
or $-$2 as above. The LAS $YJ$-only sample was uploaded to the DR4 foorprint tool to ensure that 
only sources within the footprint were considered. These sources were then cross-matched with the DR4 
catalogue, and the nearest SDSS source (for each $YJ$-only source) returned.

$YJ$-only candidates were retained if this cross-matching resulted in one of three possible criteria 
being met: (i) The nearest SDSS object was $>$\,2\arcsec~ from the LAS position, which would imply 
that the SDSS object was either a mis-match (ie the LAS source was not detected in SDSS), or that 
the SDSS counterpart was indicative of a possible high proper motion object (which could also be 
interesting). (ii) A SDSS source was found within 2\arcsec, but it was not a significant detection. 
Significant detections were defined as having a signal-to-noise $>$\,5\,$\sigma$ in any of the $ugriz$ 
bands. This criteria ensured that potentially interesting sources were not ruled out by the presence 
of an insignificant SDSS detection. (iii) A significant SDSS source was found within 2\arcsec, with 
$i-z>2$ and $z-J>$2.5. Such red optical-infrared colours are indicative of very cool objects such as 
T dwarfs (e.g. Chiu et al. 2006).

LAS images of the retained objects were then inspected, to identify and remove cross-talk artifacts 
and other spurious detections such as diffraction spikes and blended sources, and SDSS images were 
also examined to ensure the credence of the optical constraints. The final selection of $YJ$-only 
sources from the EDR included 20 candidates, all of which were DR4 non-detections. A high fraction 
(80\%) of these have J$>$19.

\section{Follow-up observations and Results}

\subsection{L dwarfs}

\begin{figure}
   \centering
   \includegraphics[width=8cm]{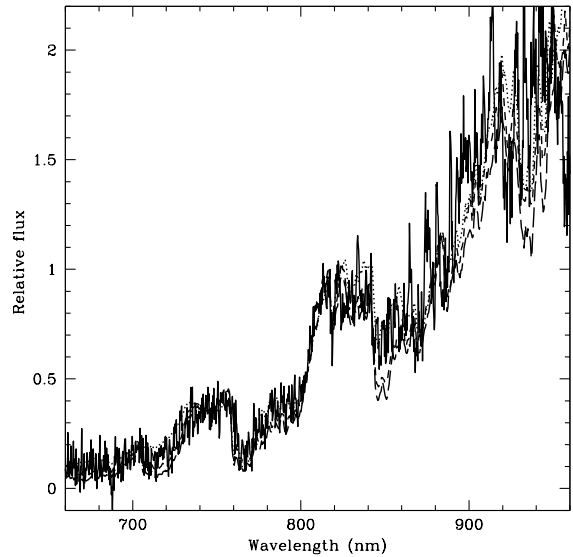}
      \caption{The spectrum of ULAS J153108+060111, an early $\sim$L0 dwarf found in the LAS EDR, is 
shown as a solid black line. Also shown for comparison are the spectra of the dL0 DENIS 0909-06 as a 
dotted line, the dM9.5 DENIS 1208+01 as a short dashed line, and the dM9 DENIS 1431-19 as a long 
dashed line. The spectrum of the dL0 \cite{mar99} is the closest match to the LAS spectrum, and has 
an estimated spectral type uncertainty of 0.5 subclasses.}
         \label{f2}
   \end{figure}

In Fig.\,\ref{f2} we show the optical spectrum of an L0 dwarf from the sample of 39 outlined in 
Sect. 2. This object (see Table 1) is the reddest of the brightest two sources in the sample (with 
$J$\,=\,15.81), and was thus selected as suitable for followup on 4m class telescopes. Spectroscopic 
observations were made with the DOLORES instrument at the 3.6\,m Telescopio Nazionale Galileo on 2006 
March 14 with the low resolution red grism yielding a resolution of 11\,\AA. The detector is a Loral 
thinned 2048\,$\times$\,2048 CCD with 15$\mu$m pixels and a 0.275\arcsec\,pix$^{-1}$ scale. Telluric 
correction and relative flux calibration was achieved using an observation of an A0 standard star made 
at a similar airmass to the target observation.

ULAS J153108+060111 yields both interesting photometric and astrometric measurements. Its SDSS/UKIDSS 
colours of $i-z$\,=\,1.95 and $z-J$\,=\,2.44 compare very well to expectations for an $\sim$\,L0 dwarf 
(\cite{kna04}, their Fig.\,3). The close agreement between the spectral type inferred from the SDSS/UKIDSS 
colours and the spectroscopically derived L0 type is extremely encouraging, and suggests that 
future searches involving LAS/SDSS cross-correlation should yield reasonably accurate spectral type 
predictions for such dwarfs. In addition, the object is bright enough to be detected in the 2MASS All-Sky 
point source catalogue, and it can be seen that its motion over a $\sim$\,5\,yr epoch difference is too 
small to be measured with any significance. This is exactly what would be expected given its estimated 
distance of 64\,pc, which was derived from the calibration of \cite{cru03}.

Further followup of the remaining L dwarf candidates from this sample has yet to be attempted, but 
could be desirable in the future, particular for the fainter and therefore more distant dwarfs. If 
improved L dwarf model atmospheres allow the $T_{\rm eff}$ of these objects to be accurately measured, 
then their distances could be inferred, and constraints placed on the disk scale-height for L dwarfs.

\subsection{T dwarfs}

This section discusses followup observations of the sample of seven 
candidates identified via $YJH$ colour cuts and $K$-band non detections. 
The most important result so far is the discovery of one certain
and one very probable T dwarf via \meth\ imaging using IRIS2
at the Anglo-Australian Telescope (AAT). The certain discovery,
ULAS J1452+0655 has been confirmed spectroscopically at Subaru 
and has a $\approx$ T4.5 spectral type. ULAS J1301+0023 has an 
almost identical \meth\ colour, and consistent $YJH$ colours, 
suggesting a very similar spectral type.

Methane observations of four candidates (see Table 1) were carried out 
on 2006 May 15 and 2006 July 8, though on both nights seeing was not 
optimal. The observations employed the \Ms\ ($1.53 - 1.63\,\mu$m) and \Ml\ 
($1.64 - 1.74\,\mu$m) methane filters installed in IRIS2. The IRIS2 
detector is a HAWAII1 1024$\times$1024 HgCdTe array with 18.5\,$\mu$m 
pixels yielding an image scale of 0.4486$\arcsec$/pix at $f$/8. Observing 
strategies, reduction and analysis procedures followed those described 
by Tinney et al. (2006), and in particular the use of 2MASS photometry 
of background objects to obtain differential \Msl\ photometry, which is 
reported in Table 1, along with the LAS merged catalogue YJH photometry 
and associated uncertainties. Two of the candidates were methane 
non-detections, and two showed significant evidence for methane 
(see Table 1).

Tinney et al. (2006) provide spectral type versus JHK colour versus
\Msl\ sequences for dwarf stars and brown dwarfs. Because the UKIDSS 
LAS adds a more sensitive Y band to its catalogue, YJH based sequences 
will be much more useful in interpreting its discoveries. This requires 
adding a Y-J versus spectral type sequence to those of Tinney et al. 
(2006), which we do in Figure 3 and Table 2. Plotted in Figure 3 is 
the synthetic Y-J photometry derived by Hewett et al. (2006) for A-T 
dwarfs on the UKIDSS photometric system, as a function of spectral 
type. Spectral type is numerically parameterised as the spectral 
sub-type, plus a constant dependent on the spectral type, where the 
constant is 0 for A, 10 for F, 20 for G, 29 for K, 35 for M, 45 for 
L and 54 for T. Note that the spectral types used by Hewett et al. 
are consistent with those used by Tinney et al. with the possible
exception of the T-types. These were therefore checked against the
compilation of hybridised T-types recently published by Burgasser et
al. (2006), (the Tinney et al. sequences are based on these types)
and where necessary the Hewett et al. types were changed to the 
Burgasser et al. types. A smoothed spline has been fitted through 
those data points to give the dwarf sequence in Table 2.

This Y-J sequence, and the J-H sequence of Tinney et al. (2006) are
plotted in Fig. 4(a) and 4(b), along with the observed photometry
of the ULAS T dwarf candidates with methane detections (large symbols) 
as well as the LAS and IRIS2 photometry of all the field objects in 
the background of each field. In both cases the UKIDSS Y-J and J-H 
colours are consistent with the relevant colour versus \Msl\ locus 
for dwarf stars, and suggest a spectral type of T3$\pm$1.

Spectroscopic observations of ULAS J1452+0655 were carried out at 
the Subaru telescope on 2006 June 16 with the CISCO (Cooled 
Infrared Spectrograph and Camera for the OH-Airglow Suppressor) 
instrument and the JH grism. The resultant wavelength coverage is 
$1.1 - 1.8\,\mu$m at a dispersion of 8.6\,$\AA$/pix with a 0.55$\arcsec$ 
wide slit. The detector was a HAWAII1 array. The integration 
time used was 30min (300s $\times$ 6 frames) and the seeing was 
measured at 0.5$\arcsec$ in the H band. An F5 telluric standard 
was observed at a similar airmass for calibration.

The resultant spectrum for ULAS J1452+0655 is plotted in Fig. 5,
{\em (solid centre line)} along with T2, T4.5 and T7 comparison spectra.
Based on these spectra we assign a spectral type of T4.5. Measurement 
of the H$_2$O-H and \meth-H indices of Burgasser et al. (2006) yield 
0.439 and 0.625 (respectively), suggesting spectral types of T3.5 and 
T3.5-T4. These measurements, are consistent with the T3$\pm$1 derived 
from \meth\ photometry and the T4.5 derived from inspection of the H 
spectrum in Fig. 5.

We give precedence to the spectral type derived by direct comparison 
with the T4.5 comparison spectrum (2MASSJ0559-1404) and adopt T4.5
type for ULAS J1452+0655. The other comparison spectra in Fig. 5 certainly
indicate that ULAS J1452+0655 cannot be as early as T2, nor as late
as T7. Given the agreement between the direct spectroscopic comparison, 
and the results given by spectral indices and methane imaging, we believe 
that the adopted spectral type is accurate to one subclass.

For a spectral type of T4.5 we adopt an absolute magnitude of M$_J$=14.1 
(Knapp et al. 2004) yielding a distance of 83pc for ULASJ1452+0655. This 
is the first demonstration that the UKIDSS LAS will deliver its stated 
potential to find Tdwarfs out to beyond 50pc. The second methane detection, 
ULASJ1301+0023, has a fainter apparent magnitude. If it has a similar 
spectral type, as seems likely, it lies at a greater distance. 
ULASJ1452+0655 is potentially therefore one of the most distant 
spectroscopically confirmed Tdwarfs yet found, standing comparison with 
other distant T dwarfs; the T6 SOri\,70 at a distance of at least 75-100pc 
and possibly as distant as 400pc (Zapatero Osorio et al. 2002); the T6 
NTTDF1205-0744 at 90pc (Cuby et al. 1999); the T3 IfA0230-Z1 at 45pc 
(Liu et al.(2002). Finally, we note that based on the IMF of Chabrier (2002, 
2003) which reproduces presently observed counts (Chabrier2005), we expect 
0.001BD/pc$^3$, and 5--6 T dwarfs in the LAS EDR, assuming a depth of J=19. 
This is reasonably consistent with the discovery of two T dwarfs from a 
sample of seven candidates in which only 5 have had follow-up observations 
sufficient to confirm or deny their T-dwarf status. Followup of the two 
additional candidates will be attempted in the future.

\begin{figure}
   \centering
   \includegraphics[width=8cm]{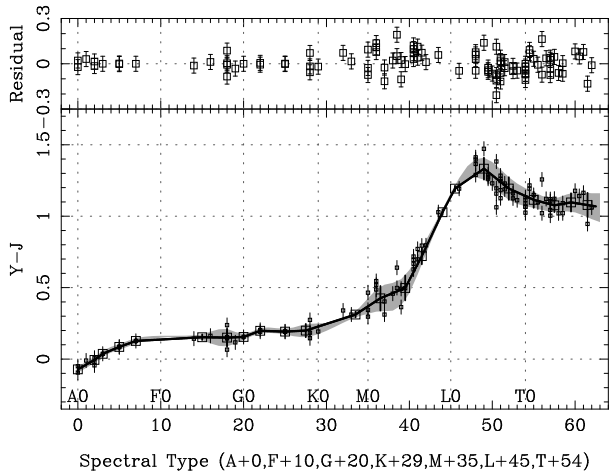}
       \caption{Y-J (UKIDSS) as a function of A-T spectral
type, following Tinney et al. (2006). Y-J synthetic photometry and spectral 
types from Hewett et al. (2006), supplemented by T-dwarf types from Burgasser 
et al. (2006). These data have been binned by spectral type {\em (large 
squares)}, which have themselves been fitted with a cubic-spline interpolating 
function {\em (solid line)}. The shaded region shows the 1-$\sigma$ rms scatter 
about the binning. These Tracks are summarised in Table 2.}
       \label{f3}
\end{figure}

\begin{table}
\caption{Y-J (UKIDSS) Dwarf Sequence based on Burgasser et al. (2006) spectral types.}
\label{t2}      
\centering          
\begin{tabular}{ll}
\hline\hline       
SpT & Y-J(UKIDSS)\\  
\hline
A0 & -0.072 \\
A5 & 0.085 \\
F0 & 0.157 \\
F5 & 0.153 \\
G0 & 0.155 \\
G5 & 0.193 \\
K0 & 0.212 \\
K3 & 0.268 \\
M0 & 0.371 \\
M2 & 0.428 \\
M4 & 0.467 \\
M6 & 0.657 \\
M8 & 0.910 \\
L0 & 1.144 \\
L2 & 1.305 \\
L4 & 1.333 \\
L6 & 1.243 \\
L8 & 1.157 \\
T0 & 1.133 \\
T2 & 1.093 \\
T4 & 1.081 \\
T6 & 1.096 \\
T8 & 1.070 \\
\hline                  
\end{tabular}
\end{table} 

\begin{figure}
   \centering
   \includegraphics[width=8cm]{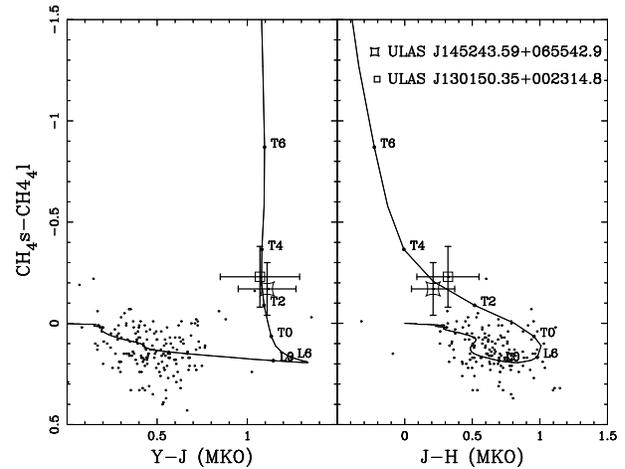}
      \caption{IRIS2 CH4 and ULAS photometry for ULAS\,J145243.59+065542.9
and ULAS\,J130150.35+002314.8 and the objects in the same IRIS2 field as these
targets (small points). The solid line is the dwarf locus in Y-J and J-H versus 
CH4 from Table 2 and Tinney et al. (2006).}
      \label{f4}
\end{figure}

\begin{figure}
   \centering
   \includegraphics[width=8cm]{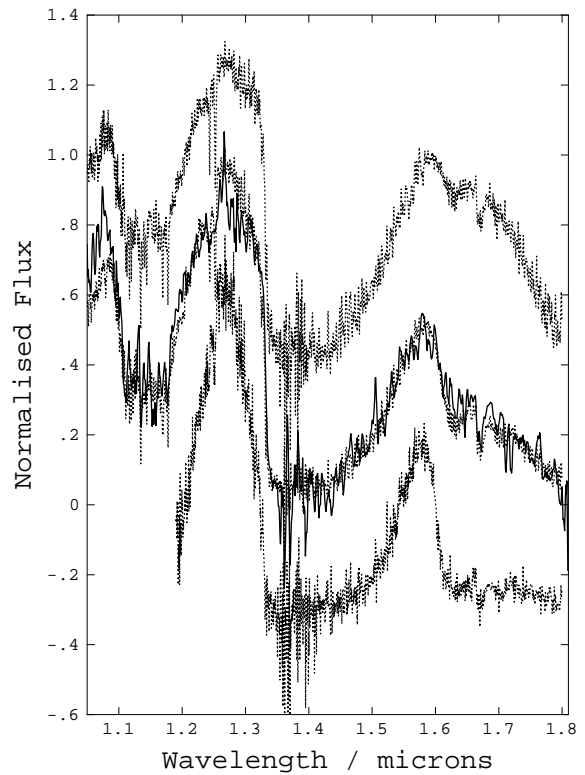}
      \caption{Subaru/CISCO spectrum of ULAS\,J\,1452+0655 (solid line, centre). Over-plotted 
on this (dotted line) is the T4.5 spectrum of 2MASSJ0559$-$1404 from \cite{cus05}. Also shown 
are the T2 spectrum of SDSS\,1254$-$0122 from \cite{cus05} and the T7 spectrum of 2MASS\,0348$-$6022 
from \cite{bur06}, shifted by +0.3 and $-$0.3 continuum units respectively.}
      \label{f5}
\end{figure}

\subsection{$YJ$-only search}

The colour space probed by the $YJ$-only search will naturally select some faint ($J>$19) late 
M dwarfs (cf Table 2), some early L dwarfs, and the majority of early-mid T dwarfs, as well as 
the sought after late T and even cooler dwarfs. It is thus important to place further constraints 
on spectral type via photometric followup observations, before spectroscopy on 8m class telescopes 
can be attempted. Optical $z$-band measurements can be used to rule out late M and early L dwarfs, 
and the latest T and cooler candidates can be confidently identified by obtaining deeper $J$- and 
$H$-band imaging to accurately measure the $J-H$ colour, as well as performing methane imaging (in 
the manner reported here). The sample of 20 $YJ$-only objects from the EDR are to be followed up 
with both optical and near infrared photometry in this way, so as to identify the most promising 
candidates for spectroscopy.

\section{Conclusions}

   \begin{itemize}
      \item We have spectroscopically confirmed a new L0 dwarf in the UKIDSS Large Area Survey 
Early Data Release using LAS and SDSS photometry. The spectral type inferred from the UKIDSS/SDSS 
photometry is in good agreement with the spectroscopic observations.
      \item We have also spectroscopically confirmed a T4.5 dwarf in the EDR, ULAS\,J\,1452+0655, 
as well as another object ULAS\,J\,1301+0023, which likely has a similar spectral type (estimated 
from our methane imaging). The discovery of two T dwarfs from 27$^2$degs of UKIDSS is in reasonable 
agreement with expectations of the surface density of T dwarfs on the sky.
      \item We have used our knowledge of LAS sensitivities and late T dwarf colours 
to identified a sample of objects that should probe the LAS for late T dwarfs (and potentially even 
cooler objects) down to the J-band limit. By selecting sources that are undetected in both the $H$- 
and $K$-bands, this search probes a significantly larger volume than comparable searches limited 
by $H$-band sensitivity.
      \item Our findings, from this small area of sky, show that the UKIDSS LAS is already performing 
up to expectation, and shows excellent promise for the future in fulfilling one of its main science 
drivers of finding unprecedentedly large numbers of brown dwarfs. 
   \end{itemize}

\begin{acknowledgements}
      TRK acknowledges funding from the UK PPARC in the context of the UKIDSS project. The authors 
are mostly members of the Cool Dwarf Science Working group (CDSWG) consortium. SKL is supported by the 
Gemini Observatory, which is operated by the Association  of Universities for Research in Astronomy, 
Inc., on behalf of the  international Gemini partnership of Argentina, Australia, Brazil, Canada, 
Chile, the United Kingdom, and the United States of America.

\end{acknowledgements}


\begin{thebibliography}{}
\bibitem[Burgasser et al. (2006)]{bur06} Burgasser, A.J., Geballe, T.R., Leggett, S.K., Kirkpatrick, 
  J.D. \& Golimowski, D.A., 2006, ApJ, 637, 1067
\bibitem[Chabrier (2002]{cha02} Chabrier, G., 2002, ApJ, 567, 304
\bibitem[2003)]{cha03} Chabrier, G., 2003, PASP, 115, 763
\bibitem[Chabrier 2005]{cha05} Chabrier, G., 2005, in ``The Initial Mass Function 50 years later'', 
  eds. Corbelli, E., Palla, F.\& Zinnecker, H., Ap\&SS Library vol. 327, Springer, Dordrecht, p. 41
\bibitem[Chiu et al. (2006)]{chiu06} Chiu, K., Fan, X., Leggett, S.K., Golimowski, D.A., Zheng, W., 
  Geballe, T.R., Schneider, D.P., Brinkmann, J., 2006, AJ, 131, 2722
\bibitem[Cruz et al. (2003)]{cru03} Cruz, K.L., Reid, I.N., Liebert, J., Kirkpatrick, J.D., \& 
  Lowrance, P.J., 2003, AJ, 126, 2421
\bibitem[Cuby et al. (1999)]{cub99} Cuby, J.G., Saracco, P., Moorwood, A.F.M., et al., 1999, A\&A, 349, L41
\bibitem[Cushing et al. (2005)]{cus05} Cushing, M.C., Rayner, J.T., \& Vacca, W.D., 2005, ApJ, 623, 1115
\bibitem[Deacon \& Hambly (2006)]{dea06} Deacon, N.R. \& Hambly N.C. 2006, MNRAS, in press, astro-ph/0607305
\bibitem[Dye et al. (2006)]{dye06} Dye, S., Warren, S.J., Hambly, N.C., et al. 2006, MNRAS, in press, 
  astro-ph/0603608
\bibitem[Hewett et al. (2006)]{hew06} Hewett, P.C., Warren, S.J., Leggett, S.K. \& Hodgkin, S.T. 2006, 
  MNRAS, 367, 454
\bibitem[Knapp et al. 2004]{kna04} Knapp, G.R., Leggett, S.K., Fan, X., et al. 2004, AJ, 127, 3553
\bibitem[Lawrence et al. (2006)]{law06} Lawrence, A., Warren, S.J., Almaini, O., et al. 2006, MNRAS, 
  in press, astro-ph/0604426
\bibitem[Leggett et al. 2005]{leg05} Leggett, S.K., Allard, F., Burgasser, A.J., Jones, H.R.A., Marley, 
  M.S. \& Tsuji, T., 2005, in Proc. 13th Cool Stars Workshop, ed. F. Favata et al. (EAS SP; Noordwijk: 
  ESA, in press (astro-ph/0409389)
\bibitem[Liu et al. (2002)]{liu02} Liu, M.C., Wainscoat, R., Mart\'{i}n, E.L., Barris, B. \& Tonry, J., 
  2002, ApJ, 568, L107 
\bibitem[Mart\'{i}n et al. (1999)]{mar99} Mart\'{i}n, E.L., Delfosse, X., Basri, G., et al., 1999, 
  AJ, 118, 2466
\bibitem[Tinney et al. (2005)]{tin05} Tinney, C.G., Burgasser, A.J., Kirkpatrick, J.D. \& McElwain, 
  M.W., 2005, AJ, 130, 2326
\bibitem[Zapatero Osorio et al. (2002)]{zap02} Zapatero Osorio, M.R., B\'{e}jar, V.J.S., Mart\'{i}n, 
  E.L., et al., 2002, ApJ, 578, 536
\end{thebibliography}
\end{document}